\documentclass[showkeys,pra,tightenlines,12pt]{revtex4-1}

\usepackage{amssymb}
\usepackage{graphicx}

\def\tr{\textrm{Tr}}

\def\ket#1{\vert #1 \rangle}

\def\ketbra#1#2{\vert #1 \rangle \langle #2 \vert}

\begin{document}

\title{Performance and error modeling of Deutsch's algorithm in IBM Q}

\author{Efrain Buksman}
\email{buksman@ort.edu.uy}
\affiliation{Facultad de Ingenier\'ia, Universidad ORT Uruguay, Uruguay.}

\author{Andr\'e L. Fonseca de Oliveira}
\email{fonseca@ort.edu.uy}
\affiliation{Facultad de Ingenier\'ia, Universidad ORT Uruguay, Uruguay.}

\author{Carolina Allende}
\email{allende@ort.edu.uy}
\affiliation{Facultad de Ingenier\'ia, Universidad ORT Uruguay, Uruguay.}

\begin{abstract}
The performance of quantum computers today can be studied by analyzing the effect of errors in the result of simple 
quantum algorithms. The modeling and characterization of these errors is relevant to correct them, for example, 
with quantum correcting codes. In this article we characterize the error of the five qubits quantum computer ibmqx4 
(IBM Q), using a Deutsch algorithm and modeling the error by Generalized Amplitude Damping (GAD) and a unitary 
misalignment operation. 
\end{abstract}


\maketitle

\section{Introduction}

The idea of using the quantum mechanics as a calculation tool arose from the pioneering works 
of Feynman and Deutsch \cite{Feynman_82, Deutsch_85} in the 80s. It is based on the use of properties as 
superposition and entanglement for the realization of computational task. This requires a quantum computer that works 
at the microscopic level. This way, a quantum computer could solve certain known problems, more efficiently than 
can be done with classical computers\cite{Raz_2018}, enabling advances in cryptography, drug research and development, 
faster data analysis, and improve artificial intelligence \cite{Dunjko_2018}.

Major companies such as Google, Intel, Microsoft and IBM, among others, have embarked on the building of quantum computers, 
being able to handle up to a several tens of qubits today. In particular,  the IBM Q machine used here, 
is a scalable quantum computer, based on superconducting technology that has the advantage of being open access
through the internet \cite{IBMQ}.

Several algorithms based on quantum advantage have been proposed; among the most important are the Grover
algorithm and the Shor factoring algorithm. The Grover algorithm\cite{Grover_1996} is a search algorithm of an element
in a disordered base. The known classical algorithms are of order $O(N)$, while the quantum algorithm allows to 
determine with high probability the desired element with an order $O(\sqrt{N})$. The Shor factoring algorithm 
\cite{Shor_1995} reduces the calculation time from a sub-exponential order in classical computing to a polynomial 
order in quantum computers.

However, the fact  that quantum systems cannot be completely isolated from the environment, together with 
systematic imperfections on gate applications, inexact state preparations, and inaccurate measurements,
induces errors in any quantum computational task \cite{Devitt_2013,Otten_2019}. While there is a fault-tolerant methods 
based on the correction of errors below a certain threshold \cite{Gottesman_2005}, these methods are very expensive 
in computational resources. In other hand, the development of new tools in order to characterize and analyze the 
effect and propagation of quantum errors is an important issue in order to try correct them with the least possible 
complexity. 

In this article we analyze the error propagation in the Deutsch algorithm (DA), a special case of the general 
Deutsch-Jozsa algorithm \cite{Deutsch_1992}, implemented in the quantum computer ibmqx4 (IBM Q). 
Some authors have implemented this algorithm in the IBM Q computer (for example to solve a 
learning parity problem \cite{Ferrari_2018}, \cite{Riste_2017}) but only some work has been done modeling the error 
of this technology\cite{Harper_2019}. The mixed state resulting from the algorithm is determined, and the error is 
characterized using an isotropic index \cite{Fonseca_2017}, and is then modeled by a standard Generalized 
Amplitude Damping error (GAD) and a misalignment unitary error model(MA). 

The public interested in research in the area of quantum algorithms, must be aware of the 
need to identified and model the errors, to ultimately correct and/or control them to obtain a satisfactory result. 

\section{Quantum computing overview}

\subsection{States}
The qubit, the unit of quantum computation, denoted by $\ket{\psi}$, is analog to a bit in a standard computation, 
and it is represented by a unitary vector, called a \emph{pure state}. For example, for one qubit state, the vector 
$\ket{\psi}=\alpha \ket{0} +\beta \ket{1}$, is a linear superposition of vectors $\ket{0}$ and $\ket{1}$, where $\alpha$ 
and $\beta$ are complex numbers that meet ${\vert\alpha\vert}^{2}+ {\vert\beta\vert}^{2}=1$. The vectors $\ket{0}$ 
and $\ket{1}$ are defined in the computational base 
as \cite{Nielsen_2000A}
\begin{equation}
\vert 0 \rangle = \left[{
\begin{array}{c} 
1 \\ 0 
\end{array}} \right] ,
\vert 1 \rangle = \left[
{\begin{array}{c} 0 \\ 1
\end{array}} \right] .
\end{equation}
A pure state of several qubits, is expressed as the linear combination on complex field of Kronecker product \cite{Henderson_1980}, 
of states of one qubit, denoted here by $\otimes$. For example for two qubits, any pure state is represented by a vector
$\vert \psi \rangle= \alpha \vert 0 \rangle \otimes \vert 0 \rangle
+\beta \vert 0 \rangle \otimes \vert 1 \rangle
+\gamma  \vert 1 \rangle \otimes \vert 0 \rangle
+\delta \vert 1 \rangle \otimes \vert 1 \rangle$, where
$\alpha, \beta, \gamma, \delta$ are complex numbers and the vector has unitary norm.
\begin{equation}
\vert \psi \rangle =
 \left[{\begin{array}{c} 
\alpha \\ \beta \\ \gamma \\ \delta 
\end{array}} \right] .
\end{equation}

\subsection{Evolution of states}
The evolution of a quantum closed system is a reversible process, and can be represented by a unitary transformation $U$,
on the state $\ket{\psi}$, $U \ket{\psi}=e^{i H \Delta t} \ket{\psi}$ where $H$ is the
Hamiltonian of the physical system and $\Delta t$ the time duration of the process \cite{Nielsen_2000A}. 
This representation is especially useful, since you can see the unitary transformations as quantum gates, 
similar to the role of classic gates in classical computation. Some transformations commonly used in quantum 
circuits are shown in the Table (\ref{Tab_gates}).
\begin{table}
\caption{\label{Tab_gates} Common Unitary matrices.}
\footnotesize
\rm
\begin{tabular*}{\textwidth}{@{}l*{15}{@{\extracolsep{0pt plus12pt}}l}}
\hline
Gate name &Matrix representation\\
\hline
\\
Hadamard gate&$H = \frac{1}{\sqrt{2}} 
\left[\begin{array}{cc} 
1 & 1   \\  
1 & -1 \end{array}\right]$\\
\\
Pauli $X$ gate&$
X=\sigma_X=\left[{\begin{array}{cc}
0  &  1\\
1  &  0\\
\end{array}}\right]$\\
\\
Pauli $Y$ gate&$
Y=\sigma_Y=\left[ {\begin{array}{cc}
0  &  -i\\
i &  0\\
\end{array}}\right]$\\
\\
Pauli $Z$ gate&$
Z=\sigma_Z=\left[ {\begin{array}{cc}
1  &  0\\
0 &  -1\\
\end{array}}\right]$\\
\\
$Cnot$ gate&$Cnot=\left[{\begin{array}{cccc}
1 & 0 & 0 & 0 \\
0 & 1 & 0 & 0 \\
0 & 0 & 0 & 1 \\
0 & 0 & 1 & 0
\end{array}}\right]$\\ 
\\
\hline
\end{tabular*}
\end{table}

\begin{figure}[ht]
\centering
\includegraphics[width=0.55\columnwidth]{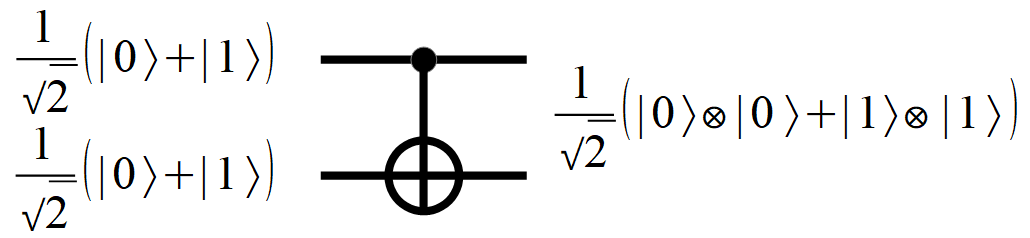}
\caption{\label{Fig_CNOT} Cnot gate aplied to state $\vert \psi \rangle $}
\end{figure}

The entanglement of a composed state (more than one qubit), is a unique feature of quantum states, that has no 
analogy on classical computing. For example, the two qubit state $GHZ$ \cite{Nielsen_2000A} is a state of 
maximum entanglement which can be generated by the application of a $Cnot$ gate (\ref{Tab_gates}) to two independent 
qubits$\ket{\psi}$
\begin{equation}
Cnot \vert \psi \rangle = Cnot \left[ \frac{1}{\sqrt{2}}\left( \vert 0 \rangle + \vert 1 \rangle \right) \otimes  \vert 0 \rangle \right]
=\frac{1}{\sqrt{2}} \left( \vert 0 \rangle \otimes \vert 0 \rangle + \vert 1 \rangle \otimes \vert 1 \rangle  \right)
\label{eq_CNOT}
\end{equation} 
as illustrated in Figure \ref{Fig_CNOT}. 

\subsection{Open systems}

When we have partial information of the state, i.e. we only have information of a subsystem of the total system, 
the state must be represented by a positive definite hermitic matrix of unitary trace, called the density matrix of 
a mixed state. Any pure state can be also described by a density matrix, but the inverse is not true. For the pure state 
$\vert \psi\rangle
=\alpha \vert 0 \rangle +\beta \vert 1 \rangle$, the density matrix denoted by 
$\vert \psi \rangle \langle \psi \vert$ is 
\begin{equation}
\vert \psi \rangle \langle \psi \vert =
\left[{\begin{array}{cc} 
\alpha \alpha^* & \alpha \beta^* \\
\alpha^* \beta  & \beta \beta^*  
\end{array}} \right].
\end{equation}

The decoherence can be thought of as an unwanted interaction with the environment \cite{Alicki_2002}. When a quantum 
system is open, the interaction between the system $S$ and the environment $E$ can be represented as 
a unitary matrix of the whole system. If we only have information about the system $S$, it can no longer be described 
as a unitary vector, but can be described by a density matrix, which is the statistical average of an assembly of 
pure states. 

The quantum evolution of a closed system can still be represented by a unitary matrix. When the state is represented by 
a density matrix $\rho$, the evolution of the state is given by $\rho^{\prime}= U \rho U^\dagger$ 
\cite{Nielsen_2000A}.

\subsection{Measurements}
Unlike the reversible evolution of a closed process, quantum measurement is an irreversible process that collapses the 
quantum state. For example, for the one qubit state $\vert\psi \rangle$, the state collapses to 
$\ket{0}$ or $\ket{1}$, with probabilities ${\vert\alpha\vert}^{2}$ and ${\vert\beta\vert}^{2}$ respectively.
In general, the measurements can be represented by operators. If we restrict ourselves to projective measurements, 
a physical observable $M$, called in this context the measurement base, can be described by the projectors $P_i$, 
generated by the eigenvectors of $M$,
\begin{equation}
M = \sum_{i} {\lambda_i P_i}=\sum_{i} {\lambda_i u_i^\dagger u_i} 
\label{eq_spectral}    
\end{equation}
where $\lambda_i$ and $u_i$ are the eigenvalues and eigenvectors respectively of $M$. 
Then, the probability after projective measurement of the state 
$\rho$ is \cite{Nielsen_2000A}
\begin{equation}
P(i)=\tr \left ( P_i \rho \right),
\label{eq_mesure_Prob_mixed}
\end{equation}
and the state after measurement becomes $\rho^\prime$
\begin{equation}
\rho^\prime = \sum_{i} {P_i \rho P_i^\dagger}.
\label{eq_measure_state_mixed}
\end{equation}
 
\section{Modeling quantum errors}

Every computational system is unavoidably affected by errors. In particular, the implementation of gates,
the preparation of states and the measurement, can have systematic errors \cite{Greenbaum_2017}, as well as errors due 
to the interaction with the environment, called decoherence errors \cite{Alicki_2002}.
Systematic errors could be modeled as random unitary matrices, but usually they have a preferential error direction. 
Both type of errors can modeled by \textit{Operator Sum Representation} (\cite{Kraus_1983}) and characterized using the 
isotropic index \cite {Fonseca_2017}

The decoherence can be modeled by a sum of operators (\textit{Operator Sum Representation} \cite{Kraus_1983}) applied to the state of the system, determined to 
better adjust the noise for each type of technology.

A quantum operation over a state $\rho$ denoted $\varepsilon(\rho)$ can be expressed as a function of Kraus operators 
($E_k$) as shown  
\begin{equation}
\centering 
\varepsilon\left(\rho \right) = \sum_{k} {E_k} \rho E_{k}^{\dagger}
\end{equation}
where $E_{k}$ satisfy $\sum _{k} E_{k}E_{k}^{\dagger }=I_{d}$.

\subsection{Generalized Amplitude Damping error (GAD)}
One of the standard models commonly used in the literature is a Generalized Amplitude Damping error (GAD), that could be 
interpreted as the interaction between a system and a thermal bath at fixed temperature. The model depends on two 
parameters: one related to the contact time with the thermal bath, represented as a probability of error $\gamma\in\left[ 0,1 \right]$, 
and the second related to the temperature of the thermal bath, represented by a parameter $p\in \left( \frac{1}{2}, 1 \right]$. 
For GAD error the Kraus operators are \cite{Nielsen_2000A},
\begin{eqnarray}
E_{0}=\sqrt{p} 
\left[{\begin{array}{cc}
1  &  0\\
0  &  \sqrt{1-{\gamma} }
\end{array}}\right],
E_{1}=\sqrt{p} \left[ {\begin{array}{cc}
0  &  0\\
0  &  \sqrt{\gamma } \end{array}} \right], \nonumber \\
\nonumber \\
E_{2}=\sqrt{1-p} \left[{\begin{array}{cc}
\sqrt{1- \gamma}  &  0\\
0  &  1
\end{array} }\right],
E_{3}=\sqrt{1-p}\left[{\begin{array}{cc}
0  &  0\\
\sqrt{ \gamma }  &  0
\end{array}}\right]
\label{GAD_error_model}
\end{eqnarray}

\subsection{Systematic errors}
In addition to a decoherence error, quantum computers suffer from systematic errors like 
classical machines. The error can be expressed by a rotation $G$, which could have a preferential direction in space. 
For example, a deviation $\epsilon$ by rotation in $X$ direction, can be represented by the unitary matrix $G= e^{i \epsilon X}$ 
where the resulting state due to error$\rho$ is  
\begin{equation}
\varepsilon(\rho) =G \rho G^\dagger 
\end{equation}

\subsection{Isotropic index}
To identify and characterize the errors, the isotropic index given in \cite{Fonseca_2017} is used. This index separates 
the part that cannot be corrected due to the total loss of information, called weight $w$, and the misalignment 
$A$ with respect to the expected reference state that, theoretically could be corrected.
Considering the pure reference state of $n$ qubits, $\rho_{\phi}=\ketbra{\phi}{\phi}$, and the decomposition of a
state $\rho$ after a noisy process, $\rho= p \frac{I}{2^n } +(1-p) \hat{\rho}$. The double index is defined as:
\begin{itemize}
\item  The Isotropic Alignment $A$, 
\begin{equation}
A = Fid (\hat{\rho},\rho_{\phi}) - Fid (\hat{\rho},\rho_{N\phi})
\end{equation}
where $Fid$ is the fidelity between quantum states, and $\rho_{N\phi}= \frac{I -\rho_{\phi}}{2^n -1}$ is the orthogonal 
isotropic mixed state of $\rho_\phi$.
\item  The isotropic weight $w = 2^n \lambda$,  with $\lambda$ being the smallest eigenvalue of $\rho$. 
\end{itemize}

The alignment $A$ takes values in the interval $[-1,1]$. When it is completely aligned
with the pure reference state $\ket{\phi}$ $A = 1$ and, when it is completely misaligned, $A = -1$. 
The weight take values in the interval $[0,1]$. When the state $\rho$ is pure $w = 0$ (the inverse is true only for one 
qubit state) and for a state without any information, i.e. maximum mixedness, $w = 1$.  

\section{Error analysis on Deutsch Algorithm}
\subsection{Deutsch Algorithm}

The Deutsch quantum algorithm, one of the first proposed, solves in the framework of quantum computing
a problem that cannot be solved in a unique step of calculation using classical computation. Although the algorithm 
proposed by David Deutsch is not of immediate application, it is the basis of important algorithms based on oracles such as Grover's search of known advantage over standard computation.

\begin{figure}[ht]
\centering
\includegraphics[width=0.5\columnwidth]{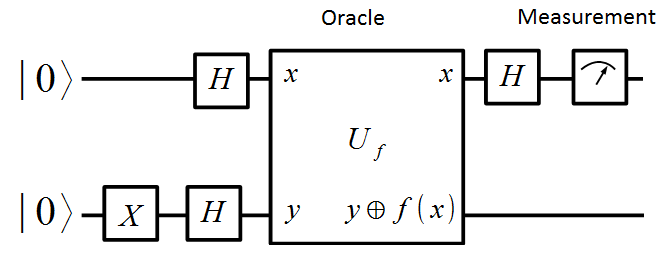}
\caption{\label{Fig_Deutsch} The Deutsch Algorithm}
\end{figure}

The idea is defined as follows; given a black box, known as the oracle, which consists of an unknown binary
function of one bit $f\left( x \right):\{0,1\} \rightarrow \{0,1\}$, the goal is to decide in a deterministic way
if the function is balanced or constant using the oracle only \textit{one time}. 
The classical algorithms need two instances of application of the oracle to solve the problem in a deterministic way, 
while quantum Deutsch algorithm uses the oracle a single instance, as long as there are no errors in the calculation process. 
The Deutsch algorithm can be summarized in the following scheme shown in the Figure \ref{Fig_Deutsch}.

Beginning with the two-qubit initial state $\vert 0 \rangle \otimes \vert 0 \rangle$, a $X$ gate is applied in the 
second qubit getting $\vert 0 \rangle \otimes\vert 1 \rangle$. After applying a Hadamard transform to each qubit, 
the result is 
$\frac{1}{2} \left(\ket{0} +\ket{1}\right) \otimes \left(\ket{0}-\ket{1} \right)$. Applying the oracle (with one 
of four possible binary functions) to the current state, and ignoring the second qubit (because at this stage the 
qubits are independent), we get
\begin{equation}
\frac{1}{\sqrt{2}} \left[\ket{0}+ \left( -1 \right) ^{f(0) \oplus f(1)}\ket{1}\right] 
\end{equation}
Finally, applying a Hadamard transform to this state we have
\begin{equation}
\frac{1}{2}\left[\left(1+ \left( -1 \right)^{f(0) \oplus f(1)} \right) \vert 0 \rangle +\left(1- \left( -1 \right)^{f(0) \oplus f(1))} \right)\vert 1 \rangle\right]
\end{equation}
Then it is concluded immediately that, the result is $0$ for constant functions, and $1$ for balanced ones.

\subsection{Deutsch Algorithm (DA) implementation on IBM Q}

\begin{figure}[ht!]		
\centering
\includegraphics[width=0.5\columnwidth]{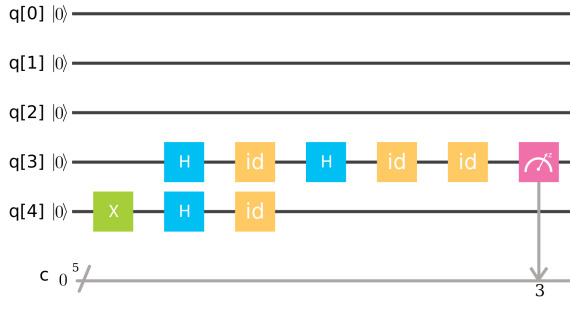}
\caption{\label{Fig_IBM_00} $DA$ IBM Q implementation, with constant zero function $f_0$.}
\end{figure}

\begin{figure}[ht!]		
\centering
\includegraphics[width=0.5\columnwidth]{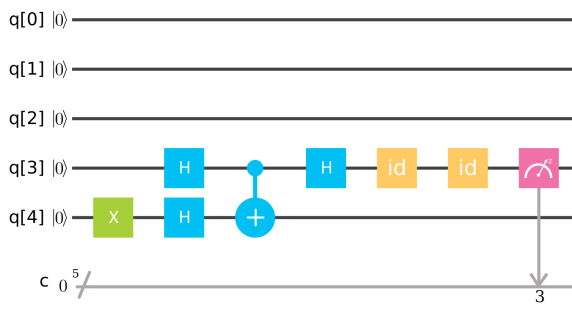}
\caption{\label{Fig_IBM_01} $DA$ IBM Q implementation, with identity function $f_{Id}$.}
\end{figure}

\begin{figure}[ht!]		
\centering
\includegraphics[width=0.5\columnwidth]{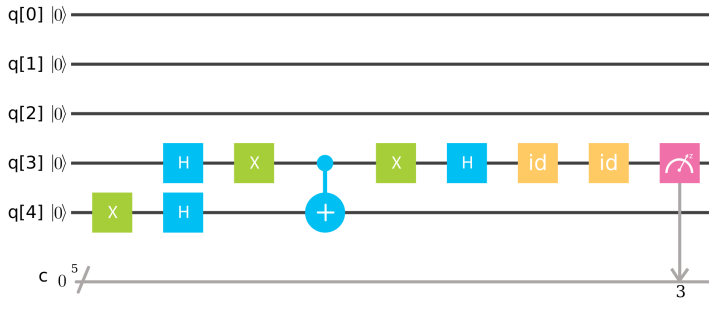}
\caption{\label{Fig_IBM_10} $DA$ IBM Q implementation, with inverse function $f_{NOT}$.}
\end{figure}

\begin{figure}[ht!]		
\centering
\includegraphics[width=0.5\columnwidth]{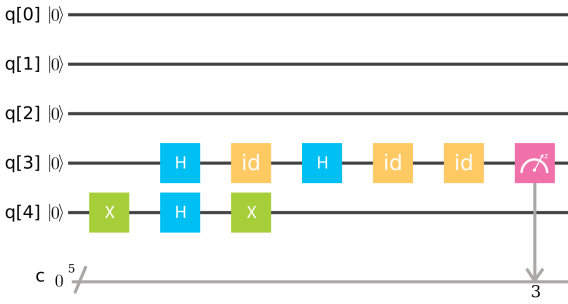}
\caption{\label{Fig_IBM_11} $DA$ IBM Q implementation, with constant one function $f_{1}$.}
\end{figure}

In order to analyze the performance of the algorithm in the IBM Q machine, the four possible functions of 
the oracle have to be implemented, i.e, $f_{0}$ \textit{constant zero}, $f_{1}$\textit{constant one},
$f_{Id}$ \textit{identity} and $f_{NOT}$ \textit{inverse function}. 
Each of these functions are implemented with the previous quantum gates Table (\ref{Tab_gates}), and are shown in Figures
(\ref{Fig_IBM_00}, \ref{Fig_IBM_01}, \ref{Fig_IBM_10}, \ref{Fig_IBM_11}).

To study the effect of the error on the resolution of the problem, the measurement of the third qubit is made (q[3]).
The result in an ideal case without error, should be 0 if the function is constant and 1 if the binary function is 
balanced. However, the actual result is affected due to several sources of error, which distort it with respect to the 
ideal case. To analyze this distortion, statistical experiments were carried out, that allow us to determine in an 
approximate way the representative assemble of the final state, i.e., the final mixed state denoted by $\rho$.

\subsection{Quantum state tomography (QST)}

To find experimentally the density matrix of a state, a method called quantum state tomography is used \cite{Nielsen_2000A}. 
Measurements must be made in the three bases (axes) of the space, $X, Y$ and $Z$, to recover the density matrix state. 
The state $\rho$ is given by \cite{Nielsen_2000A}
\begin{equation}
\rho =\frac{1}{2}\left( I_d +\tr(\rho X) X +\tr(\rho Y) Y+\tr(\rho Z) Z \right).
 \label{eq_QST}
\end{equation}
where $\tr(\rho X),\tr(\rho Y),\tr(\rho Z)$, are obtained, approaching the expected value by the statistical average. For 
example, by spectral decomposition $Z = (+1) P_0+(-1)P_1$, then, $\tr(\rho Z) = \tr(\rho P_0)- \tr(\rho P_1)$, 
that by eq. (\ref{eq_mesure_Prob_mixed}), $\tr(\rho Z)= P(0)-P(1)$, where $P(0)$ and $P(1)$ are the probabilities of measuring $0$ and $1$ respectively. Similar calculations were done
for $X$ and $Y$ operators. 

\begin{figure}[ht]		
\centering
\includegraphics[width=0.5\columnwidth]{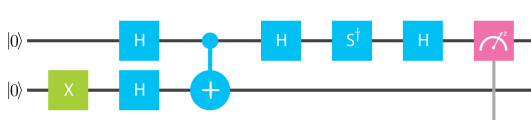}
\caption{\label{Fig_Deutsch_01_Base_Y} $DA$ IBM Q implementation, for $f_{Id}$ in $Y$ basis.}
\end{figure}

The IBM Q computer, can measure only in the canonical base ($Z$). Then, to measure on another 
base, the last two identities in figures (\ref{Fig_IBM_00},\ref{Fig_IBM_01},\ref{Fig_IBM_10} and \ref{Fig_IBM_11}), 
must be replaced. For example, to measure in the base $X$, a rotation of a $q[3]$ qubit must be made, using 
in place of $I_d \otimes I_d$, the matrix $H \otimes I_d$, and to measure in the $Y$ basis, the identities must be 
replaced by $S^\dagger \otimes H $, where $S^\dagger=\left[ {\begin{array}{cc}
1  &  0\\
0 &  -i \end{array}} \right]$, as shown in Figure \ref{Fig_Deutsch_01_Base_Y}.

For each of the four possible binary functions, the experiments were performed $8192$ times (in IBM Q computer), and the resulting density 
matrices and probabilities of success are obtained and shown in Table \ref{Tab_1}.

\begin{table}
\caption{\label{Tab_1} Probability of success for each binary function.}
\footnotesize
\rm
\begin{tabular*}{\textwidth}{@{}l*{15}{@{\extracolsep{0pt plus12pt}}l}}
\hline
Binary function&$ \ \ $Resulting matrix&Ideal result&Probability\\
\hline
\\
$f_{0}$&$\left[{\begin{array}{cc}
0.9491 & 0.0462 - 0.0664 i\\  
0.0462 + 0.0664 i &  0.0509\\
\end{array}}\right]$&$0$&$0.9491$\\
\\
$f_{I_d}$&$\left[{\begin{array}{cc}
0.1497 & 0.0270 - 0.0160 i\\  
0.0270 + 0.0160 i & 0.8503\\
\end{array}}\right]$&$1$&$0.8503$\\
\\
$f_{NOT}$&$\left[{\begin{array}{cc}
0.1748   & 0.1079 - 0.0576 i \\ 
0.1079 + 0.0576 i &   0.8252 \\ 
\end{array}}\right]$&$1$&$0.8252$\\
\\
$f_{1}$&$\left[{\begin{array}{cc}
0.9495 & 0.0510 - 0.0645i\\  
0.0510 + 0.0645i  &  0.0505\\
\end{array}}\right]$&$0$&$0.9495$\\
\\
\hline
\end{tabular*}
\end{table}

\subsection{Modeling GAD error in IBM Q }

After a statistical analysis of experimental data, it was determined that the model that best suited this algorithm 
and quantum machine (IBM Q) is a Generalized Amplitude Damping (GAD) error model [15]. Running a numerical simulation 
of this error and comparing with experimental data, the parameters that best adjust to the Eq. (\ref{GAD_error_model})
for the four functions at the same time are:
\begin{equation}
\gamma = 0.1947, p = 0.7761.
\label{optimal_GAD_param}
\end{equation}

\subsection{The Misalignment error model (MA)}
The isotropic index (\cite {Fonseca_2017}) shows that there is some misalignment for each of the 4 functions, 
with respect to the ideal result (see Table \ref {Tab_2}). 
As a result of this, we propose a method that consists of a unitary transformation applied to three qubits ( 
two qubits of the algorithm and one auxiliary one) that quantifies the misalignment of the state for both results, 0 and 1.

\begin{figure}[ht]
\centering
\includegraphics[width=0.55\columnwidth]{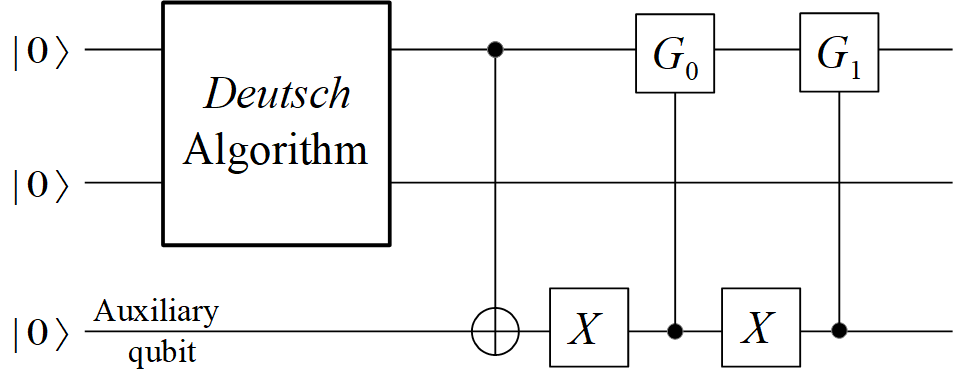}
\caption{\label{MA} The Misalignment error model(MA)}
\end{figure}

Since the result of the algorithm is a priori unknown, the model proposed applies two different unitary operators 
depending on the result: for $0$ it applies a $G_{0}$ gate and for $1$ it applies a $G_{1 }$ gate.
 These conditional operations that correct the misalignment, are given by
\begin{eqnarray}
G_{0}= \left[{\begin{array}{cc}
 0.996   &  0.053 + 0.069 i\\
0.090  &  -0.594 + 0.799 i\\
\end{array}}\right], \nonumber \\
\nonumber \\ 
G_{1}= \left[{\begin{array}{cc}
 0.994   &  - 0.097 + 0.055 i \\
- 0.111   &  -0.863 + 0.493 i  \\
\end{array}}\right]
\label{eq_giros}
\end{eqnarray}
where the conditional operation represented in Figure \ref{MA}.

The result of the model,quantified as the Fidelity between
the IBM Q experimental result and the simulated one, is shown in the Table (\ref{Tab_2}). 
\begin{table}
\centering
\caption{\label{Tab_2}Double Isotropic Index (Weight and Misalignment) and Fidelity between 
the IBM Q experimental result and the simulated error model.}
\footnotesize
\rm
\begin{tabular*}{\textwidth}{@{}l*{15}{@{\extracolsep{0pt plus12pt}}l}}
\hline
Binary function&Ideal result&Weight(w)&Misalignment(A)&Fidelity\\
\hline
$f_{0}$&$0$&0.0873&0.9070&$0.9999$\\
$f_{Id}$&$1$&0.2965&0.9544&$0.9976$\\
$f_{NOT}$&$1$&0.3051&0.8049&$0.9996$\\
$f_1$&$0$&0.0862&0.9056&$0.9999$\\
\hline
\end{tabular*}
\end{table}

\section{Results and Conclusions}

In this work we have modeled the propagation of errors in a real quantum machine, such as IBM Q (ibmqx4).
The resulting mixed states of a quantum Deutsch algorithm, are found by means of a QST method.
Through the characterization using an isotropic index, the error has been modeled, in which the loss of 
information is given by a GAD error model, and the misaligned part by means of unitary conditional 
matrix (MA error model).

As shown in Table (\ref{Tab_2}), the largest source of error is the decoherence, unlike the Grover algorithm, in which the misalignment 
is the most relevant as shown in \cite{Cohn_2016}.

The proposed error model fits very well with the experimental results, 
and can be the first step for the future correction of systematic errors in quantum systems.

\bibliographystyle{unsrt}
\bibliography{2019_RMF}

\end{document}